\documentclass[apjl]{emulateapj}
\usepackage{amssymb,amsmath,ulem}
\def\kms{km s$^{-1}$}
\def\cc{${\rm cm^{-3}}$}

\def\htwo{{\rm H_2}}
\def\um{\rm \mu m}

\shorttitle{Mass growth in MYSO fed by companion} \shortauthors{Chen
et al.}

\begin{document}
\title{Growth of a massive young stellar object fed by a gas flow
from a companion gas clump}
\author{Xi Chen\altaffilmark{1,2}, Zhiyuan Ren\altaffilmark{2,3}, Qizhou Zhang\altaffilmark{4}, Zhiqiang Shen\altaffilmark{1,2}, Keping Qiu\altaffilmark{5,6}}

\altaffiltext{1}{Shanghai Astronomical Observatories, Chinese Academy of
    Science, Nandan Rd 80, Shanghai, China}

\altaffiltext{2}{Key Laboratory of Radio Astronomy, Chinese Academy
of Sciences, Nanjing, JiangSu 210008, China}

\altaffiltext{3}{National Astronomical Observatories, Chinese
Academy of Science, Chaoyang District Datun Rd A20, Beijing, China}

\altaffiltext{4}{Harvard-Smithsonian Center for Astrophysics, 60
Garden Street, Cambridge, MA 02138, USA}

\altaffiltext{5}{School of Astronomy and Space Science, Nanjing
University, 163 Xianlin Avenue, Nanjing 210023, China}

\altaffiltext{6}{Laboratory of Modern Astronomy and Astrophysics
(Nanjing University), Ministry of Education, Nanjing 210023, China}


\begin{abstract}
We present a Submillimeter Array (SMA) observation towards the young
massive double-core system G350.69-0.49. This system consists of a
northeast (NE) diffuse gas Bubble and a southwest (SW) massive young
stellar object (MYSO), both clearly seen in the \textit{Spitzer}
images. The SMA observations reveal a gas flow between the NE Bubble
and the SW MYSO in a broad velocity range from 5 to 30 \kms {\rm
with respect to the system velocity}. The gas flow is well confined
within the interval between the two objects, and traces a
significant mass transfer from the NE gas Bubble to the SW massive
core. The transfer flow can supply the material accreted onto the SW
MYSO at a rate of $4.2\times 10^{-4}~M_\odot$ year$^{-1}$. The whole
system therefore suggests a mode for the mass growth in MYSO from a
gas transfer flow launched from its companion gas clump, {\rm
despite that the driving mechanism of the transfer flow is not yet
fully determined from the current data.}
\end{abstract}


\section{Introduction}   
Massive stars (O and B stars) contribute to the important feedback
initially on the star cluster, and ultimately drive the overall
evolution of the host galaxy through their strong outflows, stellar
winds and ionizing radiations (Kennicutt 2005). Due to short
Kelvin-Helmholtz time scale, the massive young stellar objects
(MYSOs) will experience a drastic bloating phase, during which the
stellar radius expands for $\sim30$ times larger and then rapidly
contract back to form the zero-age main sequence star (Behrend \&
Maeder 2001; Hosokawa \& Omukai 2009; Kuiper \& Yorke 2013). During
the entire process of massive star formation, how to maintain the
stable accretion to form the observed high-mass stars is thus a
major question to be concerned.

The mechanism for mass growth of massive forming star is still in
debate (see review papers, e.g., Zinnecker \& Yorke 2007; Tan et al.
2014). Different scenarios have been proposed to explain their high
masses, such as stellar collisions and mergers in very dense systems
(Bonnell et al. 1998), monolithic collapse like in low-mass star
formation (Yorke \& Sonnhalter 2002; McKee \& Tan 2003), and
competitive accretion in a proto-cluster environment (Bonnell et al.
2001, 2004; Bonnell \& Bate 2006). Accretion disks are expected at
small scales in the both scenarios, i.e., the monolithic core
collapse and competitive accretion. Recently the field of
theoretical understanding of accretion in massive star formation has
made clear process. However, two competing theories for accretion
are currently in conflict with each other: the formation of the most
massive stars via radiative Rayleigh-Taylor unstable outflows
(Krumholz et al. 2009, Rosen et al. 2016) and via disk-mediated
accretion (Nakano 1989, Yorke \& Bodenheimer 1999, Yorke \&
Sonnhalter 2002, Kuiper et al. 2010, 2011). Both scenarios solve the
radiation-pressure problem of spherically symmetric accretion flows
via an anisotropy in the thermal radiation field. The latter disk
accretion scenario has been observationally supported by e.g.
Johnston et al. (2015), while the radiative Rayleigh-Taylor
instability scenario is supported by Kumar (2013).

On larger scales, the star-forming core should have a sufficient
mass storage to feed the central star and should not be dissipated
by the stellar emission and outflow. One possibility is that the
cores obtain mass from its surrounding cores and/or the natal gas
clump. A representative model for this process is competitive
accretion (e.g., Bonnell \& Bate 2006). However, till now the
external gas supply is not fully confirmed in observations. This
should be mainly because the external gas flow into the core cannot
be so easily identified. In several ideal cases, prominent
converging flows into the hub or dense center of the filamentary
structures have been observed (e.g. Kirk et al. 2013; Peretto et al.
2013), which strongly suggests gas inflow from the surrounding
extended gas structures that is feeding the central young stars. Yet
it still calls for more extensive studies to reveal two major
properties, including 1) the specific gas motions in the
intermediate neighborhood of individual cores, whether and how the
accretion flow enters the cores; 2) the dynamical cause of the
inflow, whether it was driven by cloud collision, magnetic field or
purely due to the gravitational collapse.

In this paper, we present the observational results towards a
massive double-core system (G350.69-0.49, G350.69 hereafter), which
for the first time, exhibits an evident mass transfer flow launched
from the one core to supply the mass growth of its companion core
which is a MYSO. The double-core system shows extended shock-excited
4.5 $\mu$m emission (Extended Green Object; EGO) identified from the
Spitzer GLIMPSE II survey (Chen et al. 2013). In Section 2 we
described the observation and data reduction for the observational
data. In Section 3 we presented the overall structure of dust and
gas distribution for this source. The kinematic features and their
possible origins are more specifically discussed in Sections 4 and
5. A summary is given in Section 6.

\section{Observation and Data Reduction} 
The SMA\footnote{The Submillimeter Array is a joint project between
the Smithsonian Astrophysical Observatory and the Academia Sinica
Institute of Astronomy and Astrophysics, and is funded by the
Smithsonian Institution and the Academia Sinica.} observations of
G350.69 were carried out on 2014 May 5th in its compact array
configuration. The calibration of the time dependent antenna gains
was performed by frequent observations of quasars 1700-261 and
1744-312. The bandpass response was calibrated using the standard
calibrator quasar 3C279; and the absolute flux density was scaled by
comparing with the modelled ones of Neptune and Titan. The two 4-GHz
sidebands were centered at 219 and 229 GHz, respectively. The
channel width is 0.812 MHz, corresponding to a velocity resolution
of 1.1 km s$^{-1}$ at 230 GHz. The on-source integration time was
4.3 hours. The visibility data were calibrated using the IDL
superset MIR\footnote{see
http://www.cfa.harvard.edu/cqi/mircook.html}. The imaging and
analysis were carried out in
MIRIAD\footnote{http://www.cfa.harvard.edu/sma/miriad,
http://www.astro.umd.edu/$\sim$teuben/miriad}. The synthesized beam
size, i.e., the angular resolution of the image data is
$5.1''\times2.9''$ with a position angle of $7^\circ$ northwest and
the rms noise is 50 mJy beam$^{-1}$ per-channel for the molecular
line data, and 3 mJy beam$^{-1}$ for the 1.3 mm continuum.

The continuum emissions of G350.69 from mid- to far-infrared were
also acquired, including the Spitzer/IRAC
images\footnote{\scriptsize from the Archive of the Spitzer Enhanced
Imaging Products (SEIP), see
http://sha.ipac.caltech.edu/applications/Spitzer/SHA/} and the
Herschel PACS 70, and 100 $\um$, SPIRE 250, 350, and 500 $\um$
images \footnote{\scriptsize Herschel is an ESA space observatory
with science instruments provided by European-led Principal
Investigator consortia and with important participation from NASA.
http://www.cosmos.esa.int/web/herschel/science-archive}.

\section{Overall morphology}
\subsection{Spitzer and Herschel images}
The \textit{Spitzer} image of three-color composite at three IRAC
bands for the source G350.69 is presented in Figure 1a. It can be
clearly seen that this source resembles a binary system, containing
a northeast (NE) diffuse object and a southwest (SW) compact object.
There is a rather strong extended 4.5 $\micron$ emission (green
feature in the three-band RGB image) in the gap between the two
middle-infrared (MIR) objects along the northeast-southwest (NE-SW)
direction. There are other two 4.5 $\micron$ emission features along
northwest-southeast (NW-SE) direction approximately perpendicular to
the above mentioned NE-SW direction. As suggested by previous works
(Cyganowski et al. 2008; Chambers et al. 2009; Chen et al. 2013),
the excessive 4.5 $\micron$ features should reveral the shock
emission as induced by the supersonic material flow along these
directions.

The SW compact object should contain an MYSO due to its association
with a 6.7 GHz class II methanol maser (Minier et al. 2003; Xu et
al. 2007), which is known to be an exclusive MYSO tracer. This
object also coincides with a point source seen in \textit{Spitzer}
24 and 8 $\micron$, suggesting an embedded protostar in a dust
envelope (Chambers et al. 2009). The NE diffuse component shown in
IRAC 8 $\micron$ emission (Figure 1a) is identified as an MIR Bubble
(Churchwell et al. 2006, 2007). Dynamically formed Bubbles with
bright MIR emission require a star or cluster with UV emission to
excite the polycyclic aromatic hydrocarbon (PAH) features in the 5.8
and 8.0 $\micron$ bands (Churchwell et al. 2006). Based on its small
angular diameter of $\sim30''$ (corresponding to 0.4 pc, at a near
kinematic distance of 2.7 kpc to this source, see Chen et al. 2013)
and non-detections of H{\sc ii} region tracers, such as the
centimeter emission and radio recombination lines (Condon et al.
2008; Anderson et al. 2011), this Bubble should be produced by a
young hot late-B (below B4) star that failed to produce a detectable
H{\sc ii} region, but is still sufficiently intensive to blow a
small dust Bubble via radiation pressure (Churchwell et al. 2007).
Alternatively, the central star may be massive pre-main-sequence
star that is in a bloating stage thus has a relatively low
temperature (Kuiper \& York 2013), thereby a low ionization rate.
But it may have intense stellar wind and/or a high total luminosity
in order to generate the dusty Bubble.

The \textit{Herschel} PACS 70 $\micron$ image is presented in Figure
1b. The 70 $\micron$ emission region is reasonably coincident with
the Bubble seen in 8 $\micron$ emission. A closer look shows that
the emission is more intense towards its southwest side. This
feature is also likely seen in the \textit{Herschel} images at
longer wavelengths despite their lower resolutions, such as the
SPIRE 500 $\micron$ images as shown in Figure 1c. It may reflect the
trend that the Bubble material is being concentrated around Core-2,
the mass concentration is consistent with $^{13}$CO $(2-1)$
emissions (see below).

\subsection{Millimeter Continuum and Gas distribution}
The integrated $^{13}$CO emission and continuum emission observed
with the SMA are also shown in Figure 1. The $^{13}$CO emission
exhibits an arc-shaped structure surrounding the edge of the IRAC 8
$\micron$ gas Bubble. Such a phenomenon has also been seen in other
MIR Bubble objects. The arc-shaped emission should represent a cold
gas-and-dust Bubble believed to surround the central hot gas (Ji et
al. 2012; Dewangan et al. 2012; Dewangan \& Ojha 2013; Xu \& Ju
2014; Yuan et al. 2014; Liu et al. 2016). The UV emission should
have largely destroyed the molecules within the MIR Bubble (with a
temperature of a few 1000 K), thus no emission of molecular gas can
be seen from it.

Two significant dust continuum cores are detected in this binary
system: one is coincident with the SW MYSO (denoted as Core-1),
another near the edge of the MIR Bubble (denoted as Core-2). Core-1
is a single, compact object.  Using a two-dimensional Gaussian
profile to fit its 1.3 mm emission region (Using the \textsf{imfit}
task in {\sc miriad}), we obtained a spatial size of $(b_{\rm
maj},b_{\rm min})=(3.3'',2.4'')$. The size of Core-2 is fitted to be
$(5.2'',4.0'')$, suggesting that it is more extended than Core-1.
The extent of the Bubble is measured from its IRAC 8 $\micron$
emission (Figure 1a). The results are shown in Table 1.

\subsection{Core masses}
The Spectral Energy Distributions (SEDs) of the cores can be
constructed from their flux densities in the Herschel bands. Based
on the radiative transfer equation, the flux density of the dust
core from a gray-body emission model (Hildebrand 1983) is
\begin{equation} 
\begin{aligned}
S_{\nu} & = \kappa_\nu B_{\nu}(T_{\rm d}) \Omega \mu m_{\rm H} N_{\rm tot} \\
        & = \frac{\kappa_{\nu} B_{\nu}(T_{\rm d}) M_{\Omega} }{D^2},
\end{aligned}
\end{equation}
wherein $S_{\nu}$ is the flux density at the frequency $\nu$.
$\Omega$ is the solid angle of the core or selected area.
$B_{\nu}(T_{\rm d})$ is the Planck function of the dust temperature
$T_{\rm d}$, $N_{\rm tot}$ is the gas column density (mostly
HI+$\htwo$), $\mu=2.33$ is the mean molecular weight (Myers 1983).
$m_{\rm H}$ is the mass of the hydrogen atom. $\kappa_\nu$ is the
dust opacity; it is assumed to be related with the frequency in the
form $\kappa_\nu=\kappa_{\rm 230 GHz}(\nu/{\rm 230 GHz})^{\beta}$.
The reference value $\kappa_{\rm 230 GHz}=0.009$ cm$^2$ g$^{-1}$, is
adopted from dust model for the grains with coagulation for $10^5$
years with accreted ice mantles at a density of $10^6$ \cc\
(Ossenkopf \& Henning 1994). $D=2.7$ kpc is the source distance
(Chen et al. 2013). We note that the near distance from the Galactic
rotation curve is adopted. Another $D=10.9$ kpc, would cause the
core masses to be unreasonably high. For example, at $D=10.9$ kpc,
Core-2 mass would be $M\simeq1400~M_\odot$, which largely exceeds
the most massive cores (e.g. $M\simeq500~M_\odot$ in Perreto et al.
2013).

The apertures to measure the flux densities are plotted in dashed
lines in Figure 2b. The Bubble is overlapped with Core-2. Assuming a
uniform brightness over the Bubble surface, $S_{\rm Core2}$ was
measured by subtracting the background emission measured from the
bubble surface away from Core-2. And for the Bubble, we have $S_{\rm
Bubble}=S_{\rm Bubble+Core2}-S_{\rm Core2}$, wherein $S_{\rm
Bubble+Core2}$ is their total flux density measured within the
largest circle shown in Figure 2b.

The SED fitting is mainly affected by the flux uncertainties, which
should be evaluated. The uncertainties mainly include three factors,
including the rms noise over the image ($\Delta S_{\rm rms}$), the
errors in flux calibration ($\Delta S_{\rm cal}$), and the low
angular resolution that blends different cores ($\Delta S_{\rm
res}$). And the total uncertainty is $\Delta S_{\rm tot}^2=\Delta
S_{\rm rms}^2+\Delta S_{\rm cal}^2+\Delta S_{\rm res}^2$. In
G350.69, the uncertainties were found to be dominated by the third
part since the cores are poorly resolved in the Herschel bands. In
order to estimate its scale, we assumed each object to originally
have a 2D Gaussian distribution with the half-maximum diameter same
as that shown in Table 1.

The model image is convolved with the beam size and regridded with
the pixel size in each band. And then we compared the flux
measurement between the modeled and the observed images, and the
difference between the two may roughly represent the uncertainty. We
note that the modeled image cannot represent the real dust
distribution but is only used to estimate the flux uncertainties. As
a result, the uncertainty scales ($\Delta S_{\rm res}/S_{\nu}$) are
from 10\% to 150\% for 70 to 500 $\micron$ bands. The flux
calibration uncertainties ($\Delta S_{\rm cal}/S_{\nu}$) are around
5\% to 10\% for the Herschel bands (see PACS and SPIRE manuals, also
described in Ren \& Li 2016, Appendix A.1 therein). The rms level is
measured to be $5-20$ mJy beam$^{-1}$ throughout the Herschel bands.
Its contribution to $\Delta S_{\rm tot}$ is much less significant
than the other two terms. The measured flux densities are shown in
Table 2.

Although the flux measurement in the SPIRE bands has large
uncertainties due to the poor resolutions (Figure 2c), it does not
largely increase the $T_{\rm d}$ uncertainty mainly because the SED
is more closely constrained by the emissions in shorter wavelengths
(PACS bands). For each body, the observed flux densities can be well
reproduced using a single $T_{\rm d}$ component. $\beta$ and $M_{\rm
70 \micron}$ are also inferred from the SED fitting. The results are
presented in Table 1.

The masses of Core-1 and Core-2 are calculated also using the 1.3 mm
emissions, which should represent the masses of the compact gas
components. The Bubble is not detected in 1.3 mm due to insufficient
uv coverage for short baselines.

The column density $N({\rm H_2})$ of the three objects at their
centers are all calculated from the 70 $\micron$  using Equation
(1). The values of Core-1 and Core-2 are also derived from the 1.3
mm intensities. And the number densities $n({\rm H_2})$ are then
derived using $n(\htwo)=N(\htwo)/\bar{d}$, where $\bar{d}$ is the
average diameter of the core that is $\bar{d}=(b_{\rm maj}+b_{\rm
min})/2$. The value should represent the average number density at
the center along the line of sight.

\section{Kinematical properties}
\subsection{Mass transfer flow}
Figure 3a and 3b show the velocity-integrated $^{12}$CO emissions in
different velocity ranges. The velocity ranges of these components
can be determined from the Position-Velocity diagrams as shown in
Figure 3c.

The PV1 direction (Figure 3c, upper panel) shows that the gas flow
is distributed in a broad velocity range, extending from $|\Delta
V|=5$ to 30 \kms\ from the central velocity ($V_{\rm sys}=-19$
\kms). The low- and high-velocity components are observed to have
distinct morphologies. The low-velocity components have $|\Delta
V|<15$ \kms\, and are distributed in both blue- and red-shifted
sides. Whereas the high-velocity component is only observed in the
blueshifted side and is less intense than the low-velocity
components. It is located between Core-1 and Core-2 and with a
velocity range of $\Delta V=10-30$ km s$^{-1}$.

The integrated emissions of the low-velocity components (Figure 3a)
are mainly extended from Core-1 to Core-2.  The components along the
PV-1 direction are roughly symmetrically distributed around Core-2.
The blueshifted emission extends to Core-2 and is almost rightly
terminated at Core-1 center. Another blueshifted emission feature is
to the east of Core-2 at offset=$(-15'',0'')$. This feature is also
seen in $^{13}$CO and C$^{18}$O and should trace the gas
condensation in the Bubble shell. Figure 3b shows a very compact
morphology of the high-velocity flow between Core-1 and Core-2.
There are no other emission features in this velocity range.

In PV2 direction there are two gas lobes symmetrically distributed
with respect to Core-1. They are $20''$ distant from Core-1 center
(Figure 3a). The two lobes also have comparable velocity ranges
($\Delta V=2$ to 12 \kms) and intensities as shown in Figure 3c
(lower panel). They should represent a bipolar outflow ejected from
Core-1. It is noticed that each outflow lobe has both red- and
blueshifted emissions. This may suggest that the outflow axis is
close to the plane of the sky, so that each lobe can exhibit
opposite gas motions as projected along the line of sight.

The SiO $(5-4)$ and H$_2$CO (3-2) emissions and PV diagrams are
shown in Figure 4. The two species both have compact emission
regions around Core-1 and Core-2 (Figure 4a), but largely differ in
their velocity distributions (Figure 4b). The SiO emission has a
large fraction extending to the blueshifted side up to $\Delta V=10$
\kms\ (Figure 4b, upper panel). Similar with the high-velocity
$^{12}$CO flow, the blueshifted SiO emission is also confined
between Core-2 and Core-1. In comparison, the H$_2$CO has only one
narrow velocity component emission around the $V_{\rm sys}$ from
Core-2 to Core-1, despite that it also has weak emissions to the
blueshifted side.

The SiO emission also suggests that the mass transfer flow is
launched from Core-2 to Core-1. At the Core-2 center (offset$=12''$
in the PV diagram, Figure 4b), the SiO emission feature continuously
extend from $V_{\rm lsr}=-18$ to $-25$ \kms. While at Core-1
(offset$=0''$) there is apparently a velocity gap between the
blueshifted emission $(-27,-21)$ \kms\ and the Core-1 emission
$(-20,-15)$ \kms. A reasonable explanation is that the mass transfer
flow is being braked at Core-1 so that the velocity distribution is
also interrupted. In fact, as seen in Figure 3c, the high-velocity
component in $^{12}$CO is also connected with Core-2, likely being
accelerated and terminated towards Core-1. The fact that mass
transfer flow is started from Core-2 and terminated at Core-1
suggests that it could supply the mass for the star formation in
Core-1.

Assuming the mass transfer flow to have a cylindrical shape that
links Core-1 and Core-2, the corresponding mass transfer rate to
Core-1 can be approximately derived using $\dot{M}_{\rm trans}=\mu
m_{\rm H_2} N_{\rm H_{2}}d_{\rm flow}\bar{v}_{\rm flow}$. Here
$N_{\rm H_{2}}$ is the gas column density in the region of the
transfer flow, $d_{\rm flow}$ is the average diameter of the flow
cross section, $\bar{v}_{\rm flow}$ is the average flow velocity. In
calculation, we adopt $N_{\rm H_2}=5\times10^{22}$ cm$^{-2}$ which
is derived from the 1.3 mm dust emission between Core-1 and Core-2
($\sim$0.014 Jy beam$^{-1}$). The flow diameter $d_{\rm flow}$ is
adopted as width of the SiO emission region (Figure 4a) deconvolved
with the beam size, that is $d=5.0''$ ($\sim$14000 AU). The flow
velocity is adopted as the mean velocity of the low-velocity
component in $^{12}$CO ($v_{\rm flow}=7$ \kms\ as measured from
Figure 3c). Based on these values, we estimate the mass transfer
rate of 4.2$\times10^{-4}$ $M_\odot$ yr$^{-1}$.

We note that the $^{12}$CO blue lobe (Figure 3a) apparently has a
larger width than the SiO emission region, namely a larger $d_{\rm
flow}$. It would imply a higher mass transfer rate if the gas traced
by the $^{12}$CO blue lobe can be all obtained by Core-1.

\subsection{Core rotation and Bipolar outflow from Core-1}
The integrated emission region and the intensity-weighted velocity field (moment-1 map) of the C$^{18}$O (2-1) are shown in Figure 5a. The PV diagrams of the C$^{18}$O and $^{13}$CO lines are shown in Figure 5b. The C$^{18}$O shows a linear velocity gradient throughout Core-1, with the radial
velocity range of $V_{\rm lsr}=(-18.5,-16)$ \kms\ through
Core-1 from its northeast to southwest. The velocity gradient should
indicate the core rotation with an average velocity of $V_{\rm lsr}=1.2$ \kms.

Comparing with Figure 3a, one can see that the two outflow lobes of
the $^{12}$CO in PV2 direction are actually aligned quite well along
the rotational axis. This suggests that Core-1 may contain a
disk-jet system, with the mid-plane roughly in parallel with the
direction from Core-2 to Core-1. The velocity at northeast edge of
Core-1 is smoothly connected with the mass transfer flow arrived on
its edge (Figure 5b, upper panel). This suggests that the mass
transfer flow may be also bringing angular momentum into Core-1,
thereby help sustain or enhance its rotation.

Assuming a $^{12}$CO abundance of $10^{-4}$ and an excitation
temperature similar with the Core-1 dust temperature (21 K, see
Table 1), the masses of the two outflow lobes were estimated to be
$M=0.5$ and 0.7 $M_\odot$, for the blue and red lobes, respectively.
From their average velocity ($\sim$ 6 km s$^{-1}$) and distance
($\sim15''$, corresponding to $4\times10^{4}$ AU) from Core-1, the
time scale of the outflow is estimated to be $t_{\rm
out}=4.8\times10^4$ year, and the outflow rate is $\dot{M}_{\rm
out}=M_{\rm out}/t_{\rm out}\simeq2\times 10^{-5}~M_\odot$
year$^{-1}$, which is much smaller than the mass transfer rate due
to the gas flow. This suggests that Core-1 should be dominated by
the transfer flow thus in a mass growth.

\section{Discussion: Origination of the transfer flow}
\subsection{Possibility: An outflow from Core-2}
The first possibility is that the transfer flow between the two
cores is a part of jet-like outflow driven by an embedded protostar
in Core-2 on the Bubble edge. It is in a morphological agreement
with the low-velocity $^{12}$CO components that have symmetric blue-
and red-shifted emission regions around Core-2 (see Figure 3a).
However, it should be questioned why the high-velocity flow (Figure
3b) does not have a redshifted counterpart on the northeast side of
Core-2. There are three possible causes for such asymmetry. First,
the high-velocity outflow might be intrinsically unipolar or has a
very weak red lobe. Second, the blue lobe might be dissipated in the
hot Bubble. Third, the high-velocity flow could be compressed by the
infalling gas into Core-2.

The second and third cases both anticipate the outflow to have a
significant interaction with the Bubble or infalling gas. Then we
would expect them to generate shocked region. However, the
redshifted CO outflow lobe is not evidently detected in either SiO
or IRAC 4.5 $\micron$ emission, suggesting that the gas interaction
between Core-2 and the Bubble is not significant, or at least weaker
than that between Core-1 and Core-2.

\subsection{Possibility: A Roche overflow}
Another possible kinematics process for the observed mass transfer
flow is that the material is the Roche overflow from the MIR Bubble
to the Core-1. This scenario is proposed mainly based on the two
distinct features which are already shown above: (1) a gas flow from
Core-2 and Core-1 with broad velocity range, (2) the rotation and
bipolar outflow in Core-1. Based on these two features, the entire
system vividly resembles a binary system with the Roche overflow
exceeding the Lagrange L1 point to accrete onto the companion star.
The likelihood of such scenario can be evaluated based on the
geometry and the core masses.

A critical position in the Roche-overflow system is the Lagrange
point L1, where the gravities from the two objects reach a balance.
If the material from one of the binary members moves over this
point, it would accrete onto the companion object if there are no
other perturbations. The distance from L1 to Core-1 center $r$ can
be estimated using the equation:
\begin{equation}
\frac{M_2}{(R-r)^2}=\frac{M_1}{r^2}+\frac{M_2}{R^2}-\frac{r(M_1+M_2)}{R^3},
\end{equation}
wherein $R$ is the distance between the two companions, $M_1$ and
$M_2$ are their masses (assuming $M_2>M_1$ without losing
generality). In G350.69, the masses are adopted to be $M_2=M_{\rm
Core2+Bubble}$ and $M_1=M_{\rm Core1}$, using the masses measured
from the 70 $\micron$ continuum (see Table 1).  The mass center of
$M_2$ is relatively uncertain, and we considered two limits that the
$M_2$ center varies from the Bubble center to the Core-2 center.

Figure 6 shows the $M_2$ position and the corresponding L1 point
range, which extends from the Bubble edge to the interval between
Core-2 and Core-1. Since the SiO and $^{13}$CO emissions are all
continuously distributed from Core-1 to Core-2, the gas flow would
obviously propagate over L1 and accretes onto Core-1.

In the classical condition of a Roche overflow, the gas motion is
determined by the equivalent gravity field in the binary system. To
examine this, we made a simple model to estimate the gas velocity
from the core masses and compare it with the observed velocity
distribution. We first calculated the gravitational potential field
due to the four gas components:

(1). The Bubble which is assumed to be concentrated within a
half-elliptical shell with number density $n_{\rm 0,Bubble}$, and
its projected image was adjusted to be best coherent with the
$^{13}$CO image.

(2,3). The dense inner components of Core-1 and Core-2, which
have uniform density and radii equal to the deconvolved radii of the
SMA 1.3 mm cores.

(4). The envelope of Core-1, which has a power-law density profile
out of $r_{\rm Core-1}$, that is $n(r)=n_{\rm 0}(r/r_{\rm
0})^{-p}~(r>r_0)$, wherein the power-law index is adopted to be
$p=-1.1$, as the average value for the massive molecular cores
(Butler \& Tan 2012). We note that the extended gas around Core-2
should be included in the Bubble shell, thus an envelope for
Core-2 was no more separately modelled.

For each component, its reference number density $n_0$ is adjusted
so that the total mass is consistent with the observed value. For
the envelope of Core-1, the mass should be $M_{\rm env}=M_{70
\micron}-M_{1.3mm}=13~M_\odot$ (see Table 1). The surface contour of
each component is shown in Figure 7a. The projected gas distribution
on the X-Y plane is shown in Figure 7b, wherein the integrated
$^{13}$CO emission is also overlaid.

In the reference frame co-rotating with the two cores, the
equivalent potential well out of the core boundaries ($|{\bf
r}|>r_{\rm core1}$ and $|{\bf r-R_{12}}|>r_{\rm core2}$) is:
\begin{equation}
\begin{aligned}
\phi_{\rm cores}({\bf r})= &-\frac{GM_1}{|{\bf r}|}-\frac{GM_2}{|{\bf r-R_{12}}|} \\
                           &-\frac{1}{2}[\Omega\times({\bf r}-\frac{M_1 {\bf R_{12}}}{M_1+M_2})]^2,
\end{aligned}
\end{equation}
wherein ${\bf R_{12}}$ is the distance vector from Core-1 to Core-2,
$\Omega$ is the angular velocity of their rotation. It is estimated
from the velocity difference between Core-1 and Core-2 and their
distances assuming an inclination angle of $45^\circ$ with respect
to the line of sight. And the masses are $M_{1,2}= M_{\rm
Core1,Core2}$ in 1.3 mm because the Equation (5) is only for the
dense cores (the extended component is modeled in Equation (6) as
following).

The second contribution is from more extended gas components,
including the Bubble and the Core-1 envelope that is not sampled
in the SMA 1.3 mm continuum. The potential field is numerically
sampled using
\begin{equation}
\phi_{\rm ext}({\bf r}) = -G \int \frac{{\rm d}M({\bf r_{\rm ext}})}{|{\bf
r-r_{\rm ext}}|}.
\end{equation}
The integration is numerically performed for the Bubble and
the envelope in the model. And the total potential is
\begin{equation}
\phi({\bf r})=\phi_{\rm cores}({\bf r})+\phi_{\rm ext}({\bf r}).
\end{equation}

For the fraction of gas that is falling into the cores originally
from Bubble center, the gas velocity purely due to $\phi({\bf r})$
would be
\begin{equation}
v_{\rm inf}({\bf r})=\sqrt{2[\phi({\bf r_0})-\phi({\bf r})]},
\end{equation}
wherein ${\bf r_0}$ is the vector distance form Core-1 center to the
Bubble center. We note that owing to the viscosity deceleration and
the projection effect, the modelled velocity field should represent
the upper limit for the observed velocity along the line of sight.
The derived velocity distribution is shown in Figure 7c. The highest
velocity is reached at Core-2 center, and Core-1 center also has a
local maximum of $v_{\rm inf}\simeq 4$ \kms, which are both within
our expectation.

Figure 7d to 7f show the observed $^{12}$CO, $^{13}$CO and SiO
velocity distribution from Core-2 to Core-1, respectively (same as
the PV diagrams shown in Figure 3 to 5). The modelled velocity
profile is overlaid in red-square line. One can see that for the
$^{13}$CO (2-1) and SiO (5-4) line, the calculated velocity profile
is reasonably consistent with the maximum value of the gas flow,
except for that a fraction of the SiO (5-4) emission is beyond the
modelled curve around Core-1. For the $^{12}$CO (2-1), the
blueshifted emission feature, in particular the high-velocity
component, largely exceeds the modelled curve. This indicates that
the high-velocity component of the flow cannot be generated solely
by the gravity of the two cores.

Besides an outflow from Core-2, the central star in the Bubble  an
also accelerate the mass transfer flow. In the case of Roche
overflow, the gas could be originally pushed towards the Roche lobe
boundary by the radiation pressure and/or stellar wind from the
central star in the Bubble. In this case, Roche lobe boundary should
be largely overlapped with the observed bubble shell (Figure 5a).
The gas would be compressed onto the shell except at the L1 point,
wherein the gas can directly move to Core-1 due to the Roche
overflow and is not accumulated therein. Therefore, the driving
force from the central star would accelerate the flow without
resistance, forming the high-velocity flow as observed. In the mean
time, the materials on the shell would slowly slide onto the L1
point and join into the transfer flow.

As a short summary for the mass transfer flow, the gravity from
Core-1 and Core-2 would be almost sufficient to pull the gas from
the Bubble and generate the low-velocity mass transfer flow. Whereas
the high-velocity flow is likely to require an additional driving
force. The outflow from Core-2 might be a possible case, while the
stellar wind and radiation from the progenitor star of the Bubble is
also a considerable factor. We note that the low-velocity flow is
much more intense than the high-velocity part and would contribute a
major fraction of the mass transfer rate.

\subsection{Contribution to the Mass Growth in Core-1}
Although we can not fully determine the initial driving mechanism
for the mass transfer flow from the current data, the observations
reveal that the Core-1 is obtaining mass via the transfer flow and
has likely formed a disk-outflow system in its accretion. In an
isolated isothermal collapsing core, the mass infall would reach a
maximum if the entire core is in a free-fall collapse, that is
$\dot{M}\approx M_{dc}/t_{\rm ff}$, wherein $t_{\rm
ff}=[3\pi/(32G\rho)]^{1/2}$ is the free-fall time, calculated to be
$t_{\rm ff}\simeq1.2\times10^5$ years. For Core-1, using the density
from the 1.3 mm continuum ($\rho=3.0\times10^{-18}$ $g$ cm$^{-3}$,
see Section 3.3 and Table 1), the mass infall rate is derived to be
5.0$\times10^{-4}$ $M_\odot$ yr$^{-1}$. This value is comparable
with the mass transfer flow ($4.2\times10^{-4}$ M$_{\odot}$
yr$^{-1}$), suggesting that infall and accretion in Core-1 can be
supplied by a large fraction from the mass transfer flow.

\section{Summary}   
In this paper we present an observational study towards the
high-mass star-forming region G350.69. The major findings are:

(1) The region contains an extended Bubble-and-shell object, and two
dense massive mass cores defined as Core-1 and Core-2. Core-1 is a
young high-mass star forming object associated with a 6.7 GHz
CH$_3$OH maser. Core-2 is located on the Bubble edge, overlapped
with its gas shell and have similar radial velocity with the
surrounding gas. It could be formed during the mass assembling
process in the shell. The overall geometry of this region is similar
with the binary-star system except for their much larger spatial
scale.

(2) A prominent gas transfer flow between Core-2 and Core-1 is
observed in several molecular lines, in particular $^{12}$CO (2-1)
and SiO (5-4). The velocity structures suggest that the flow is from
Core-2 to Core-1. The gas flow could provide a high accretion rate
of $\dot{M}_{\rm flow}=4.2\times10^{-4}~M_\odot$ year$^{-1}$ into
Core-1 which is comparable with the infall rate in a free-fall
collapse. This suggests that the mass infall and accretion onto the
central star in Core-1 could be largely sustained by the mass
transfer flow.

(3) Core-1 is rotating and launching a bipolar outflow along its
rotational axis. The outflow rate ($2\times10^{-5}$ year$^{-1}$) is
much smaller than the mass transfer rate, suggesting that Core-1 can
have a considerable mass growth. The mass transfer flow is smoothly
connected with the velocity gradient due to the Core-1 rotation.

Although we can not determine the origination of the the
material transfer flow from Core-2 to Core-1 (corresponding to a
part of outflow from an embedded protostar in Core-2, or via a Roche
overflow). All above findings support that mass transfer flow can
considerably sustain the growth of the high-mass young star embedded
in Core-1, therefore suggesting a distinct mode of the mass growth
of MYSO via a gas transfer flow launched from its companion gas
clump.

\section*{Acknowledgment}
Our results are based on the observations made using the
Submillimeter Array (SMA), which is a joint project between the
Smithsonian Astrophysical Observatory and the Academia Sinica
Institute of Astronomy and Astrophysics and is funded by the
Smithsonian Institution and the Academia Sinica. This research has
also made use of the data products from the GLIMPSE and MIPSGAL
surveys, which are legacy science programs of the \textit{Spitzer}
Space Telescope funded by the National Aeronautics and Space
Administration. This work was supported by the National Natural
Science Foundation of China (11590781, 11403041 and 11273043), the
Strategic Priority Research Program ``The Emergence of Cosmological
Structures'' of the Chinese Academy of Sciences (CAS), Grant No.
XDB09000000, the Knowledge Innovation Program of the Chinese Academy
of Sciences (Grant No. KJCX1-YW-18), the Scientific Program of
Shanghai Municipality (08DZ1160100), and Key Laboratory for Radio
Astronomy, CAS. K.Q. acknowledges the support from National Natural
Science Foundation of China (NSFC) through grants NSFC 11473011 and
NSFC 11590781.

{\small
\section{References}
Anderson, L. D., Bania, T. M., Balser, D. S., \& Rood, R. T. 2011, ApJS, 194, 32 \\
Behrend, R., \& Maeder, A. 2001, A\&A, 373, 190 \\
Bonnell, I. A., Bate, M. R., Clarke, C. J., \& Pringle, J. E. 2001, MNRAS, 323, 785\\
Bonnell, I. A., Vine, S. G., \& Bate, M. R. 2004, MNRAS, 349, 735\\
Bonnell, I. A., \& Bate, M. R. 2006, MNRAS, 370, 488\\
Butler, M. J., \& Tan, J. C. 2012, ApJ, 754, 5\\
Caselli, P., Walmsley, C. M., Zucconi, A., et al. 2002, ApJ, 565, 344\\
Chambers, E. T., Jackson, J. M., Rathborne, J. M., \& Simon, R. 2009, ApJS, 181, 360\\
Chen, X. et al. 2013b, ApJS, 206, 9\\
Chen, X., Gan, C.-G., Ellingsen, S. P., et al. 2013a, ApJS, 206, 9\\
Churchwell, E., Povich, M. S., Allen, D., et al. 2006, ApJ, 649, 759\\
Churchwell, E., Watson, D. F., Povich, M. S., et al. 2007, ApJ, 670, 428\\
Condon, J. J., Cotton, W. D., Greisen, E. W., et al. 1998, AJ, 115, 1693\\
Cyganowski, C. J., Whitney, B. A., Holden, E., et al. 2008, AJ, 136, 2391\\
Dewangan, L. K., Ojha, D. K., Anandarao, B. G., Ghosh, S. K., \& Chakraborti, S. 2012, ApJ, 756, 151\\
Dewangan, L. K., \& Ojha, D. K. 2013, MNRAS, 429, 1386\\
Frerking, M. A., Langer, W. D., \& Wilson, R. W. 1982, ApJ, 262, 590\\
Hildebrand, R. H. 1983, QJRAS, 24, 267\\
Hosokawa, T., \& Omukai, K. 2009, ApJ, 691, 823 \\
Ji, W.-G., Zhou, J.-J., Esimbek, J., et al. 2012a, A\&A, 544, A39 \\
Johnston, K., et al. 2015, ApJ, 813, 19\\
Kennicutt, R. C. 2005, in IAU Symposium, Vol. 227, Massive Star Birth: A Crossroads of Astrophysics, ed. R. Cesaroni, M. Felli, E. Churchwell, \& M. Walmsley, 3-11\\
Kirk, H., Myers, P. C., Bourke, T. L., et al. 2013, ApJ, 766, 115\\
Kirk, H., et al. 2013, ApJ, 766, 115\\
Krumholz, M., et al. 2009, Science, 323, 754\\
Kuiper, R., \& York, H., 2013, ApJ, 722, 61\\
Kuiper, R., \& Yorke, H. W. 2013, ApJ, 772, 61\\
Kuiper, R., et al. 2010, A\&A, 511, 81\\
Kuiper, R., et al. 2011, ApJ, 732, 20\\
Kumar, M. 2013, A\&A, 558, 119\\
McKee, C., \& Tan, J., 2003, ApJ, 585, 850\\
Minier, V., Ellingsen, S. P., Norris, R. P., \& Booth, R. S. 2003, A\&A, 403, 1095\\
Myers, P. C. 1983, ApJ, 270, 105\\
Nakano, T. 1989, ApJ, 345, 464\\
Ossenkopf, V., \& Henning, T. 1994, A\&A, 291, 943\\
Peretto, N. 2013, A\&A, 555, 112\\
Peretto, N., Fuller, G. A., Duarte-Cabral, A., et al. 2013, A\&A, 555, A112\\
Rosen, A., et al. 2016, MNRAS, 463, 2553\\
Tan, J., et al. 2014, in Protostars and Planets VI, University of Arizona Press, Tucson, 914 pp., p.149-172\\
Xu, J.-L., \& Ju, B.-G. 2014, A\&A, 569, A36\\
Xu, Y., Li, J. J., Hachisuka, K., et al. 2008, A\&A, 485, 729\\
Yorke, H., \& Bodenheimer, T. 1999, ApJ, 525, 330\\
Yorke, H., \& Sonnhalter, C. 2002, ApJ, 569, 846\\
Yorke, H., \& Sonnhalter, C. 2002, ApJ, 569, 846  \\
Zinnecker, H., \& Yorke, H. W. 2007, ARA\&A, 45, 481\\
}


\begin{table*}
{\scriptsize \centering
\begin{minipage}{90mm}
\caption{The physical properties of the cores.}
\begin{tabular}{lccc}
\hline\hline
Parameters                            &  Core-1              &  Core-2              &  Bubble  \\
\hline
$M_{\rm 70 \micron}(M_\odot)^{a}$     &  28                  &  47                  &  75       \\
$M_{\rm 1.3 mm}(M_\odot)$             &  15                  &  35                  &  $-$      \\
$M_{\rm C^{18}O}(M_\odot)$            &  -                   &  -                   &  70       \\
$\beta$                               &  2.3                 &  2.7                 &  2.0       \\
$T_{\rm d}({\rm K})^a$                &  21                  &  24                  &  22        \\
$b_{\rm maj}$                         &  $3.3''$             &  $5.4''$             &  $15''$   \\
$b_{\rm min}$                         &  $2.4''$             &  $4.0''$             &  $10''$   \\
$PA^c$                                &  $10^\circ$          &  $5^\circ$           &  $20^\circ$        \\
$N({\rm H_2})_{70}$(cm$^{-2}$)$^{b}$  &  $8.5\times10^{22}$  &  $3.5\times10^{23}$  &  $3.9\times10^{22}$  \\
$N({\rm H_2})_{1.3}$(cm$^{-2}$)$^{b}$ &  $1.2\times10^{23}$  &  $4.6\times10^{23}$  &  $-$  \\
$n({\rm H_2})_{70}$(cm$^{-3}$)        &  $3.5\times10^5$     &  $1.0\times10^6$     &  $1.0\times10^5$ \\
$n({\rm H_2})_{1.3}$(cm$^{-3}$)       &  $7.8\times10^5$     &  $2.5\times10^6$     &  $-$ \\
\hline
\end{tabular} \\
$a.${ The dust temperatures are obtained from SED fitting of the flux densities in Herschel
70, 160, 250, 350, and 500 $\micron$ bands.} \\
$b.${ $N({\rm H_2})$ of the Bubble is calculated from the 70 $\micron$ emission at the Bubble center.} \\
$c.${ The position angle is measured counter clockwise from north direction.}
\end{minipage}
}
\end{table*}

\begin{table*}[t]
{\scriptsize \centering
\begin{minipage}{90mm}
\caption{The flux densities of the three objects in Janskys.}
\begin{tabular}{lccc}
\hline\hline
Parameters           &  Core-1       &  Core-2      &  Bubble     \\
\hline
$F_{70}$             &  135(13)      &  56(6)       &  82  (8)        \\
$F_{160}$            &  110(13)      &  83(10)      &  147(18)        \\
$F_{250}$            &  67(24)       &  61(22)      &  73 (26)        \\
$F_{350}$            &  16(10)       &  18.2(12)    &  18.7(12)      \\
$F_{500}$            &  4.0(7)       &  9.0(12)     &  4.5(5)       \\
$F_{\rm 1.3mm}^a$    &  0.16         &  0.3         &  $<0.01$    \\
\hline
\end{tabular} \\
$a.${The Bubble is too extended to be observed in SMA 1.3 mm, thus
the emission is mainly from Core-2.}
\end{minipage}
}
\end{table*}

\clearpage



\begin{figure*}
\centering
\includegraphics[angle=0,width=0.9\textwidth]{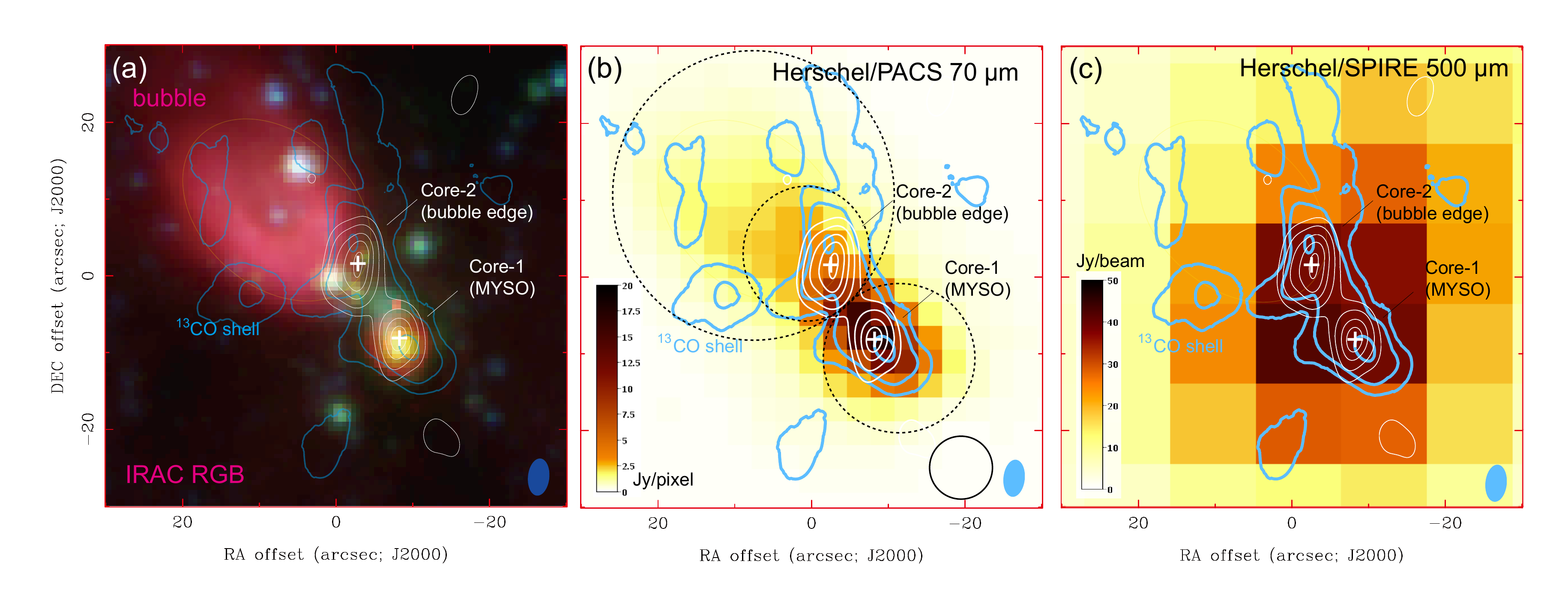} \\
\caption{\small {\bf (a)} Overall structures of the binary G350.69-0.49 from the Spitzer and SMA observations. The background is the Color-synthesized image from the Spitzer/IRAC 3.6(blue), 4.5(green), 8.0(red) $\micron$ bands. The red emission in the 8.0 $\micron$ band highlights the hot gas Bubble illuminated by the central star; the 4.5 $\micron$ emission is largely contributed by the shock excited CO and H$_2$ emission, and thus may trace the interaction between the high-velocity gas flow and the surrounding gas. The blue contours represent the velocity integrated $^{13}$CO emission. The contour levels are 4, 10, 20, 30 times 0.75 Jy beam$^{-1}$ km s$^{-1}$.The thin white contours represent the 1.3 mm continuum emission which reveals two dense dust-and-gas cores located at the Bubble edge and the MYSO, respectively.The contour levels are 4, 10, 20, 30, 40, 50 times 0.003 Jy beam$^{-1}$.  {\bf (b)} The $^{13}$CO (2-1) and 1.3 mm continuum emissions overlaid on the Herschel/PACS 70 $\micron$ continuum image (gray scale). The two dashed circles represent the apertures to measure the flux density of Core-1 and the total flux density of Core-2+Bubble. {\bf (c)} Same as (b) except for that the background image is the SPIRE 500 $\micron$ image. }    
\end{figure*}

\begin{figure*}
\centering
\includegraphics[angle=0,width=0.9\textwidth]{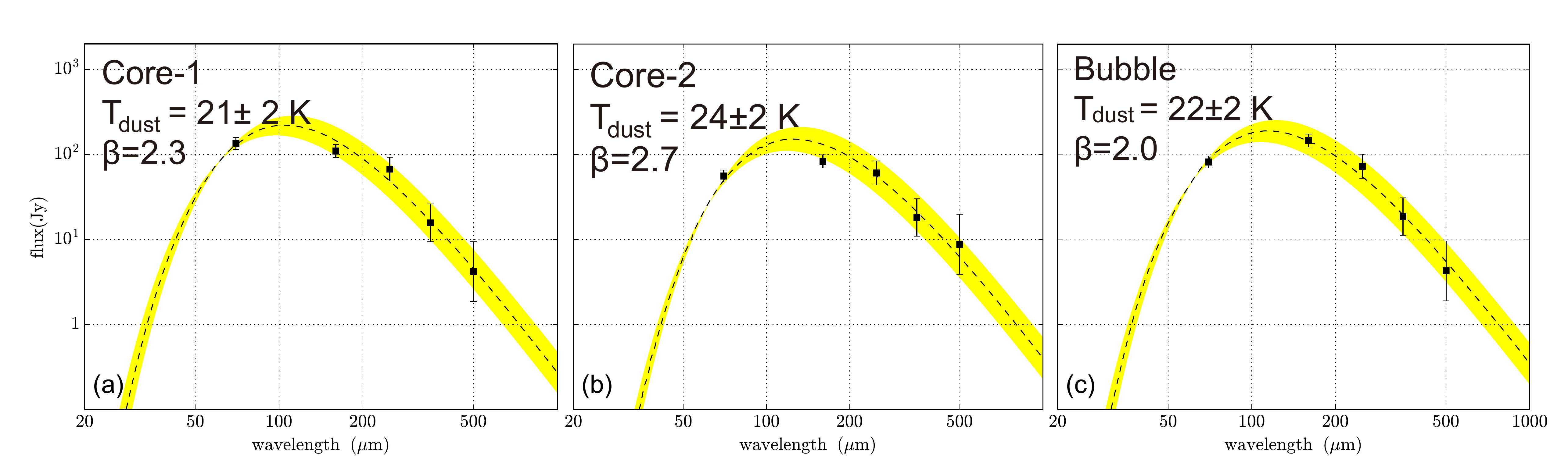} \\
\caption{\small  The spectral energy distributions (SEDs) for the three objects. The apertures for measuring the flux densities are presented in Figure 1b. The yellow-shaded area represent the variation of the best-fit SED curve due to the flux uncertainties. }    
\end{figure*}

\begin{figure*}
\centering
\includegraphics[angle=0,width=0.9\textwidth]{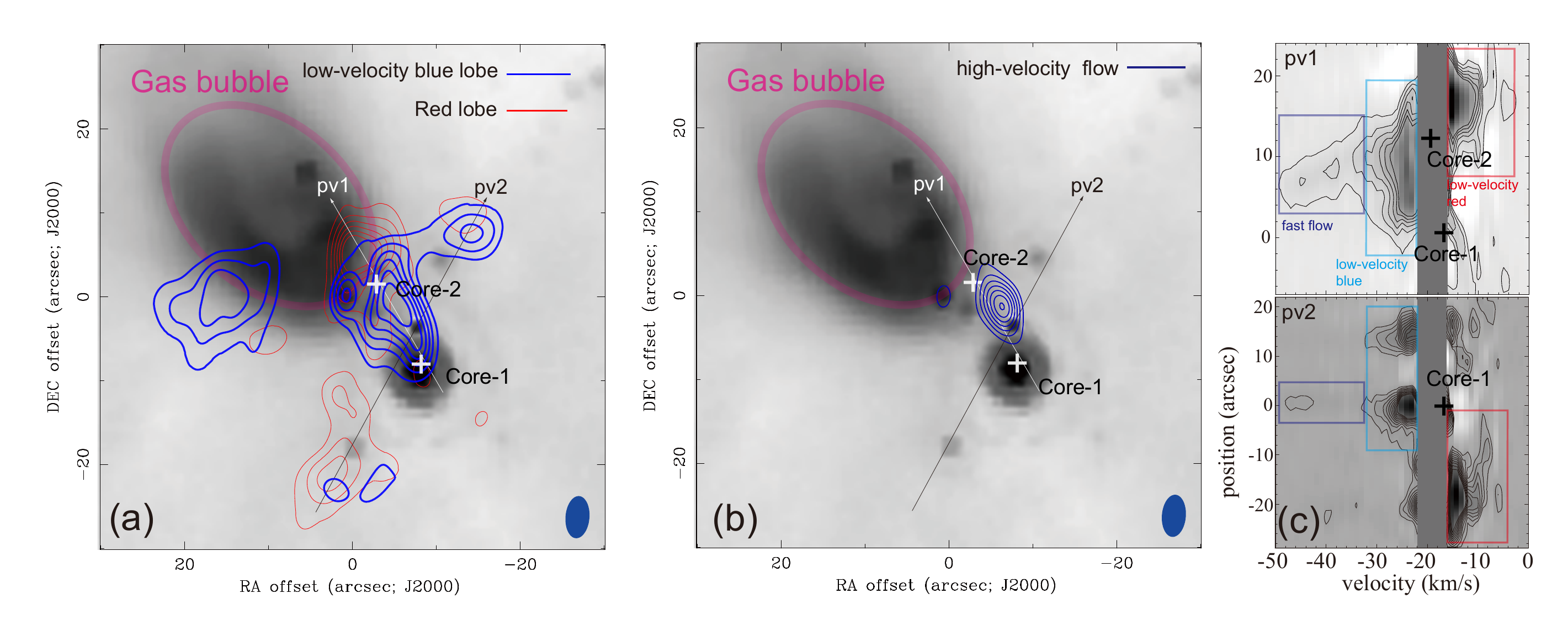} \\
\caption{\small Molecular gas kinematics of the G350.69-0.49 region from the SMA observations. {\bf (a)} The $^{12}$CO low-velocity flow components in different velocity ranges. The cyan and red represent the low-velocity flow in (-32,-22) and (-16,-4) km s$^{-1}$ ranges, respectively. And the contour levels are 4, 10, 20, 30, 40 times 0.72 Jy beam$^{-1}$ km s$^{-1}$; the white crosses represent the location of the two dense cores. {\bf (b)} The fast flow in the velocity interval of (-50,-32) km s$^{-1}$; the contour levels are 4, 6, 8, 10, 12 times 0.86 Jy beam$^{-1}$ km s$^{-1}$. {\bf (c)} The position-velocity diagrams along the two directions labelled in the main panel, with the velocity range and the bulk spatial extent of the gas flows indicated by the colored boxes. The contour levels are from 0.5 K (0.27 Jy beam$^{-1}$) in step of 1 K.}    
\end{figure*}

\begin{figure*}
\centering
\includegraphics[angle=0,width=0.7\textwidth]{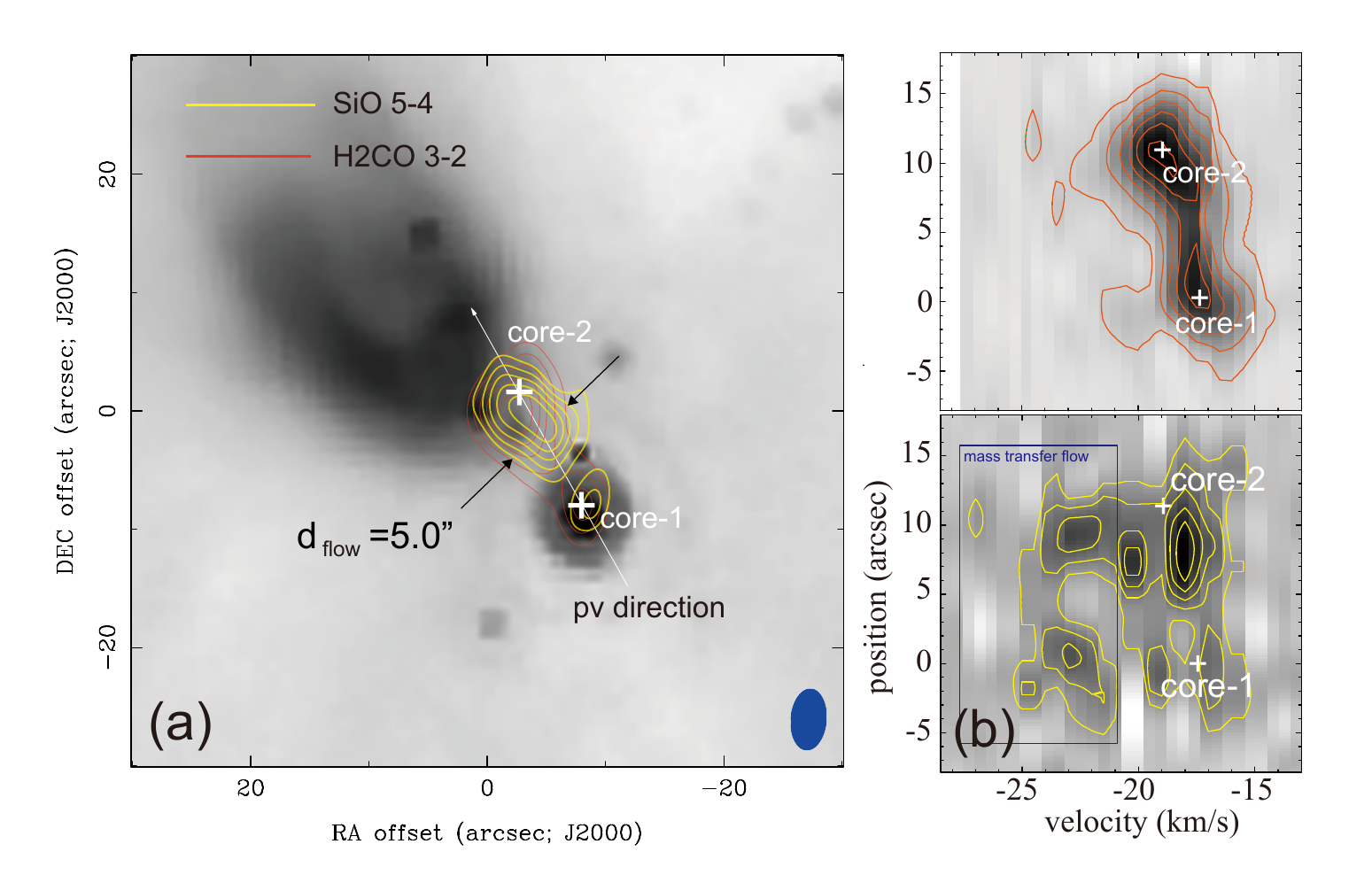} \\
\caption{\small {\rm (a)} The velocity integrated emissions of the SiO (5-4) and H$_2$CO (3-2) lines overlaid on the IRAC 8 $\micron$ image. For both species, the contours are 4, 6, 8... times the rms level (0.5 Jy beam$^{-1}$ km s$^{-1}$). The diameter of the flow cross section $(d_{\rm flow})$ is measured and labeled on the figure.  {\rm (b)} the PV diagrams of the two lines along the velocity gradient over Core-1. The contour levels are 0.2 to 1.8 K in step of 0.2 K for SiO, and 0.3 to 5.1 K in step of 0.6 K for H$_2$CO. }    
\end{figure*}

\begin{figure*}
\centering
\includegraphics[angle=0,width=0.7\textwidth]{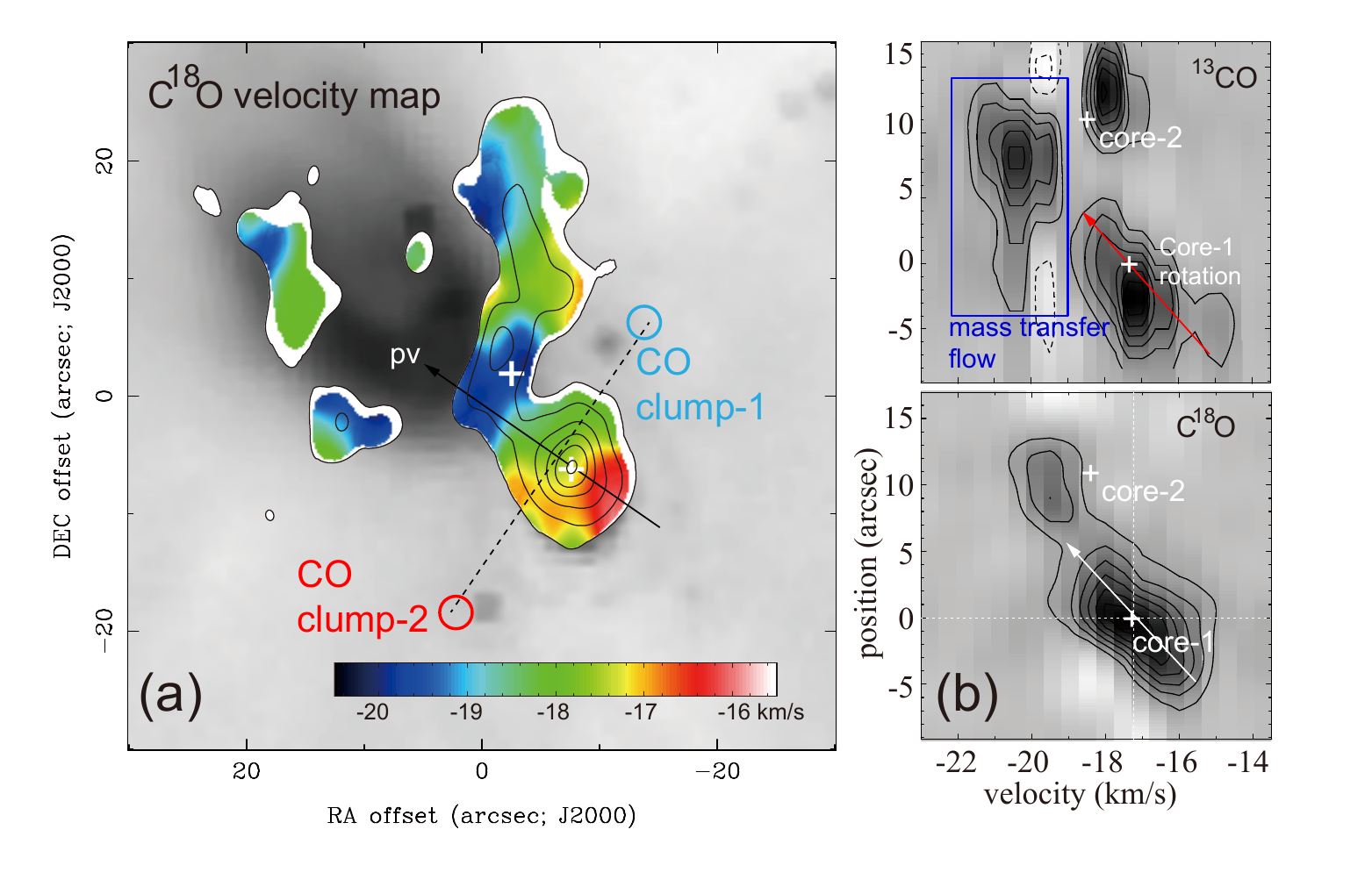} \\
\caption{\small {\rm (a)} The velocity integrated emission of C$^{18}$O (2-1) and its intensity-weighted velocity distribution (moment-1 map) overlaid on the IRAC 8 $\micron$ image. The contour levels are 4, 10, 20, 30, 40, 50 times the uncertainty level (0.5 Jy beam$^{-1}$ km s$^{-1}$). {\rm (b)} the PV diagrams of the $^{13}$CO and C$^{18}$O and (2-1) along the direction from Core-1 to Core-2. The contour levels are 0.5 to 3.0 K in step of 0.5 K for $^{13}$CO, and 0.8 to 5.3 K in step of 0.9 K for C$^{18}$O. }    
\end{figure*}
 
\begin{figure*}
\centering
\includegraphics[angle=0,width=0.5\textwidth]{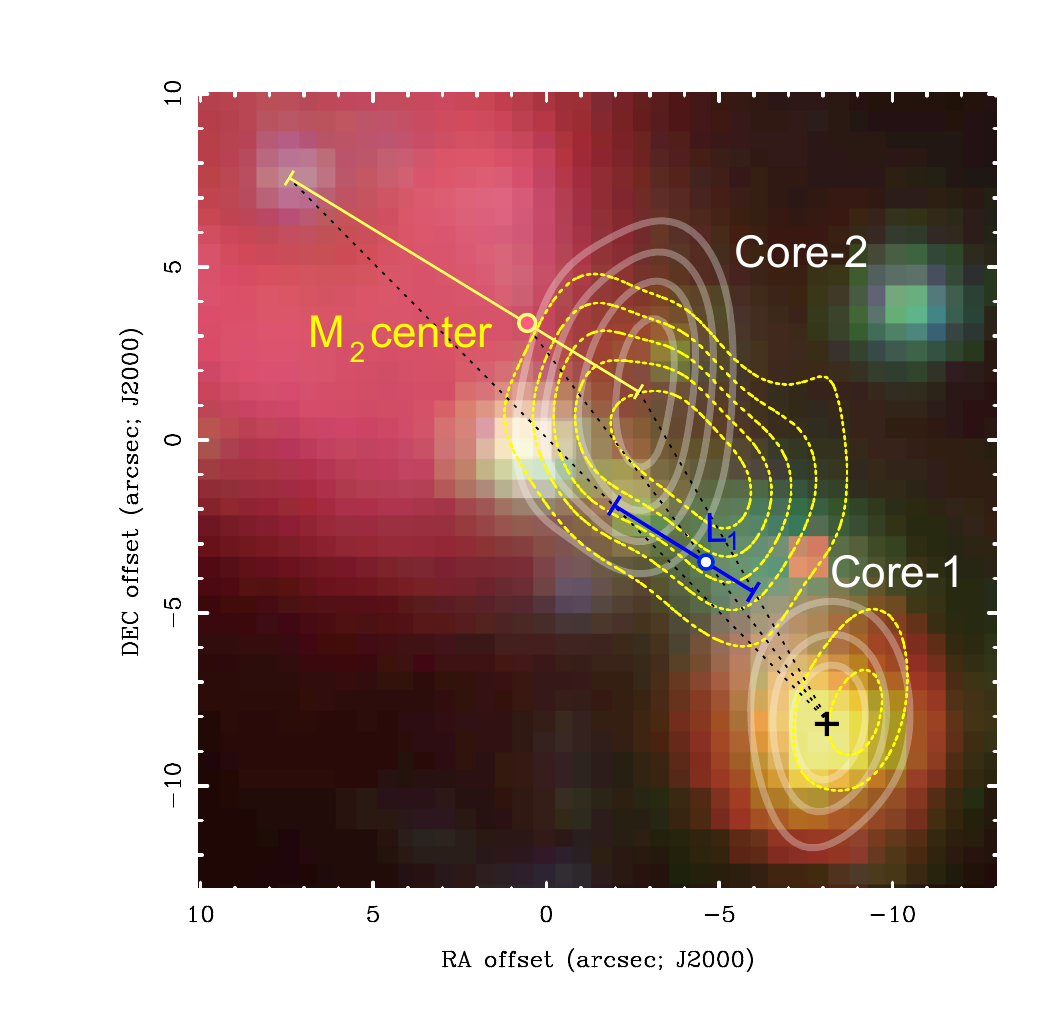} \\
\caption{\small The Lagrange L1 point between the two cores as calculated from a binary model using Equation (4). The uncertainty range is estimated by assuming the mass center of M2 (including Bubble and Core-2) varying from the Bubble center to Core-2 center, The two limits are represented by the error bars.}    
\end{figure*}

\begin{figure*}
\centering
\includegraphics[angle=0,width=0.5\textwidth]{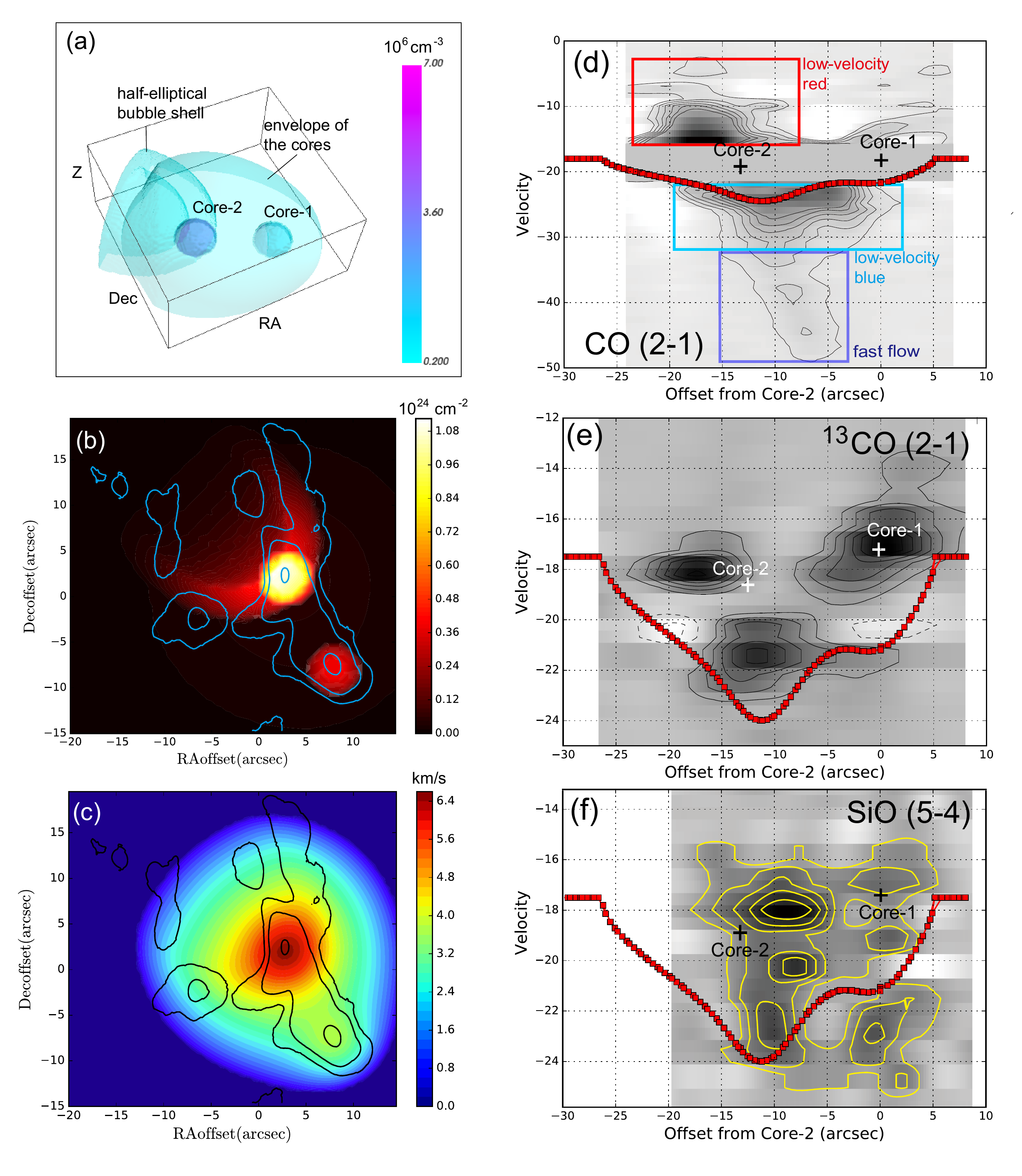} \\
\caption{\small The simplified gas structures modeling and the resultant velocity field purely due to the their overall gravity. {\bf (a)} The 3D gas components used to model the observed gas morphologies. {\bf (b)} The gas column density map overlaid on the observed $^{13}$CO image. {\bf (c)} The velocity distribution calculated from Equation (8), assuming gas is infalling from the Bubble center, representing the maximum velocity that the can be reached  {\bf (d)}-{\bf (f)} The modeled velocity distribution along the direction from Core-2 to Core-1 overlaid on the PV diagrams of $^{12}$CO, $^{13}$CO, and SiO lines, respectively. }      
\end{figure*}


\end{document}